\documentclass[a4paper]{PoS}
\usepackage{amsmath}
\usepackage{amssymb}
\usepackage{fontenc}
\usepackage{mathptmx}
\usepackage{graphicx}

\title{A resolution of the inclusive flavor-breaking sum rule $\tau$ 
$V_{us}$ puzzle}

\ShortTitle{Resolving the $\tau$ $V_{us}$ puzzle}

\author{\speaker{Kim Maltman}\\
Dept. Mathematics and Statistics, York University, Toronto, ON
Canada M3J 1P3\\
CSSM, Department of Physics, University of Adelaide, Adelaide, SA 5005 
Australia \\
E-mail: \email{kmaltman@yorku.ca}}

\author{Renwick Hudspith, Randy Lewis, Carl Wolfe\\
Dept. Physics and Astronomy, York University, Toronto, ON Canada M3J 1P3\\
E-mail: \email{renwick.james.hudspith@gmail.com, randy.lewis@yorku.ca,
wolfe@yorku.ca}}

\author{James Zanotti\\
CSSM, Department of Physics, University of Adelaide, Adelaide, SA 5005
Australia\\
E-mail: \email{james.zanotti@adelaide.edu.au}}

\abstract{A combination of continuum and lattice methods is used to
investigate systematic issues in the finite-energy-sum-rule 
determination of $V_{us}$ based on flavor-breaking 
combinations of hadronic $\tau$ decay data. Results for $V_{us}$ 
obtained using assumptions for $D>4$ OPE contributions employed in 
previous conventional implementations of this approach are shown to 
display significant unphysical dependences on the choice of sum rule 
weight, $w$, and upper limit, $s_0$, of the relevant experimental 
spectral integrals. Continuum and lattice results suggest the
necessity of a new implementation of the flavor-breaking sum rule 
approach, in which not only $\vert V_{us}\vert$, but also $D>4$ 
effective condensates are fit to data. Lattice results also provide a 
means of quantifying the truncation error for the slowly converging $D=2$ 
OPE series. The new implementation is shown to produce $\vert V_{us}\vert$ 
results free of unphysical $s_0$- and $w$-dependences and typically 
$\sim 0.0020$ higher than the (unstable) results found using the 
conventional implementation. With preliminary new experimental results 
for the $K\pi$ branching fraction, the resulting $\vert V_{us}\vert$ 
is in excellent agreement with that obtained from $K_{\ell 3}$, and 
compatible within errors with expectations from three-family unitarity. }

\FullConference{The 33rd International Symposium on Lattice Field Theory\\
		14 -18 July 2015\\
		Kobe International Conference Center, Kobe, Japan*}

\begin{document}

\section{Introduction}
The conventional $\tau$ decay determination of $\vert V_{us}\vert$ is 
based on finite-energy sum rules (FESRs) involving flavor-breaking (FB) 
combinations of inclusive hadronic $\tau$ decay data~\cite{gamizetal}. 
With $\Pi^{(J)}_{V/A;ij}(s)$ the $J=0,1$ components of the flavor 
$ij=ud,us$, vector (V) or axial vector (A) current 2-point
functions, $\rho^{(J)}_{V/A;ij}(s)$ their spectral functions, and 
$\Delta\Pi_\tau \, \equiv\,
\left[ \Pi_{V+A;ud}^{(0+1)}\, -\, \Pi_{V+A;us}^{(0+1)}\right]$,
one has
\begin{equation}
\int_0^{s_0}w(s) \Delta\rho_\tau (s)\, ds\, =\, 
-{\frac{1}{2\pi i}}\oint_{\vert
s\vert =s_0}w(s) \Delta\Pi_\tau (s)\, ds\ ,
\label{basicfesr}
\end{equation}
valid for any $s_0$ and any analytic $w(s)$. The spectral 
function, $\Delta\rho_\tau$, of $\Delta\Pi_\tau$, 
is experimentally accessible in terms of the 
differential distribution, $dR_{V/A;ij}/ds$, of the normalized ratio
$R_{V/A;ij}\, \equiv\, \Gamma [\tau^- \rightarrow \nu_\tau
\, {\rm hadrons}_{V/A;ij}\, (\gamma )]/ \Gamma [\tau^- \rightarrow
\nu_\tau e^- {\bar \nu}_e (\gamma)]$. Explicitly~\cite{tsai71}
\begin{eqnarray}
&&{\frac{dR_{V/A;ij}}{ds}}\, =\, c^{EW}_\tau \vert V_{ij}\vert^2
%{\frac{12\pi^2\vert V_{ij}\vert^2 S_{EW}}
%{m_\tau^2}}}\,
\left[ w_\tau (s ) \rho_{V/A;ij}^{(0+1)}(y_\tau )
- w_L (y_\tau )\rho_{V/A;ij}^{(0)}(s) \right]
\label{basictaudecay}\end{eqnarray}
with $y_\tau =s/m_\tau^2$, $w_\tau (y)=(1-y)^2(1+2y)$, 
$w_{L}(y)=2y(1-y)^2$, $c^{EW}_\tau$ a known constant,
and $V_{ij}$ the flavor $ij$ CKM matrix element. $\Delta\Pi_\tau$
on the RHS of Eq.~(\ref{basicfesr}) is to be treated using the OPE.

The reason for employing the $J=0+1$ FESR Eq.~(\ref{basicfesr}),
rather than the analogue involving the spectral function combination in 
Eq.~(\ref{basictaudecay}), is the very bad behavior of the integrated 
$J=0$, $D=2$ OPE series~\cite{longprob}. $J=0$ contributions to 
$dR_{V/A;ud;us}/ds$ are determined phenomenologically and subtracted, 
allowing $\rho_{V/A;ud,us}^{(0+1)}(s)$ to be obtained. The subtraction 
is dominated by the accurately known, non-chirally-suppressed 
$\pi$ and $K$ pole contributions. Continuum contributions to 
$\rho_{V/A;ud}^{(0)}$ are $\propto (m_d\mp m_u)^2$ and 
numerically negligible, while small, but not totally negligible, 
$(m_s\mp m_u)^2$-suppressed continuum $\rho_{V/A;us}^{(0)}$ 
contributions are fixed using highly constrained dispersive and 
sum rule methods~\cite{jop,mksps}. With $\vert V_{ud}\vert$ 
known~\cite{ht14}, $\Delta\rho_\tau (s)$ is expressible in terms of
experimental data and $\vert V_{us}\vert$. $\vert V_{us}\vert$
is then obtained by using the OPE for $\Delta\Pi_\tau$ on the RHS
and data on the LHS of Eq.~(\ref{basicfesr}).

Given the $J=0$-subtracted $dR^{(0+1)}_{V+A;ud,us}/ds$, it is straightforward 
to define re-weighted $J=0+1$ versions of $R_{V+A;ud,us}$, 
$R^w_{V+A;ij}(s_0)\equiv \int_0^{s_0}ds\, {\frac{w(s)}
{w_\tau (s)}}\, {\frac{dR^{(0+1)}_{V+A;ij}(s)}{ds}}$, for any
$w$ and $s_0\le m_\tau^2$. With $ \delta R^{w,OPE}_{V+A}(s_0)$
the OPE representation of $\delta R^w_{V+A}(s_0)\, =\,
{\frac{R^w_{V+A;ud}(s_0)}{\vert V_{ud}\vert^2}}
\, -\, {\frac{R^w_{V+A;us}(s_0)}{\vert V_{us}\vert^2}}$,
one then has
\begin{equation}
\vert V_{us}\vert \, =\, \sqrt{R^w_{V+A;us}(s_0)/\left[
{\frac{R^w_{V+A;ud}(s_0)}{\vert V_{ud}\vert^2}}
\, -\, \delta R^{w,OPE}_{V+A}(s_0)\right]}\ .
\label{tauvussolution}\end{equation}
The resulting $\vert V_{us}\vert$ should be independent of $w(s)$ and 
$s_0$, provided external experimental and theoretical inputs, and any 
assumptions employed in evaluating $\delta R^{w,OPE}_{V+A}(s_0)$, are 
reliable. Since integrated $D=2k+2$ OPE contributions scale as $1/s_0^k$, 
problems with assumptions about higher $D$ non-perturbative contributions
will show up as instabilities in $\vert V_{us}\vert$ as a function of $s_0$.

\begin{figure}[ht]
\includegraphics[width=.38\textwidth,angle=270]
{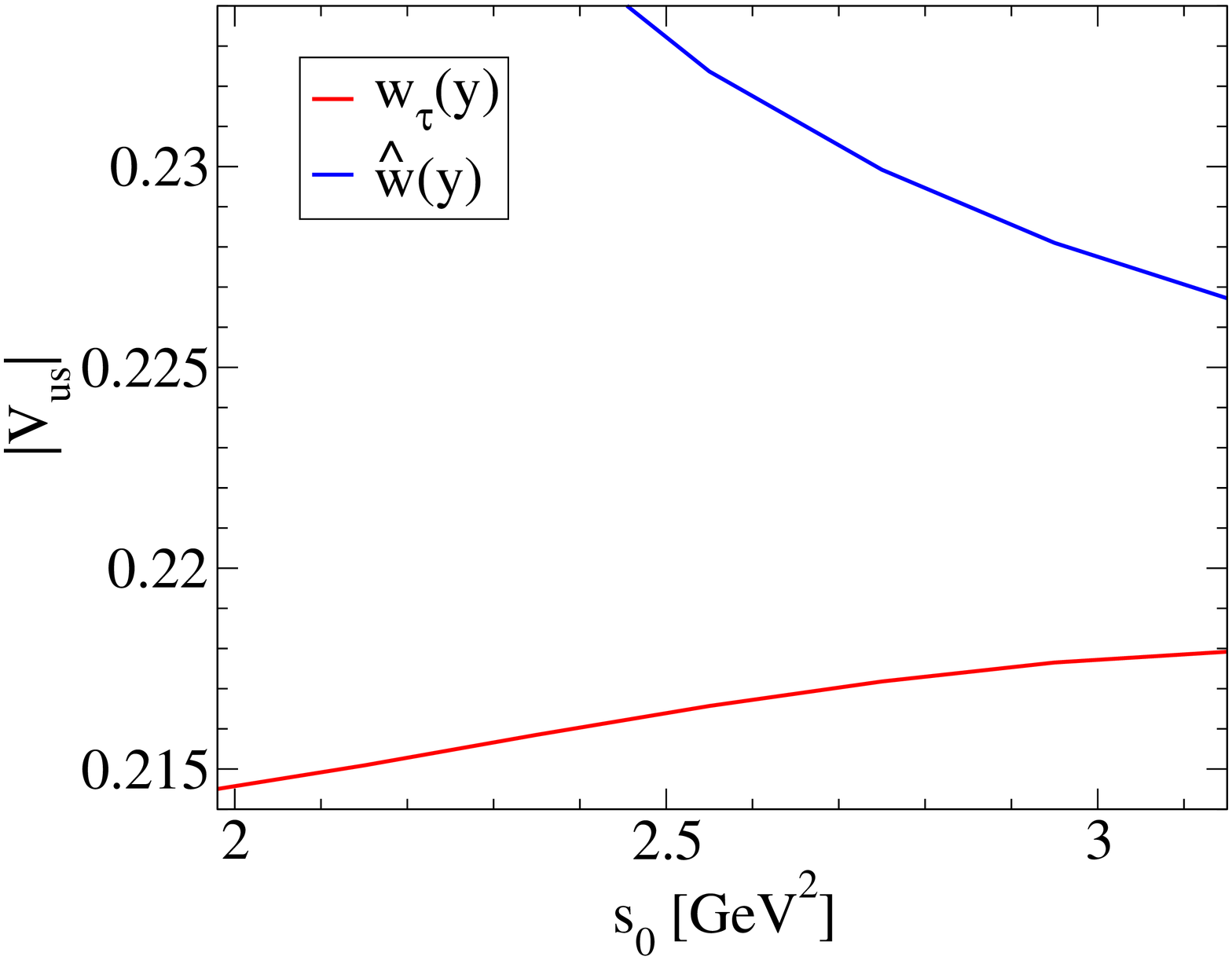}
\ 
\includegraphics[width=.38\textwidth,angle=270]
{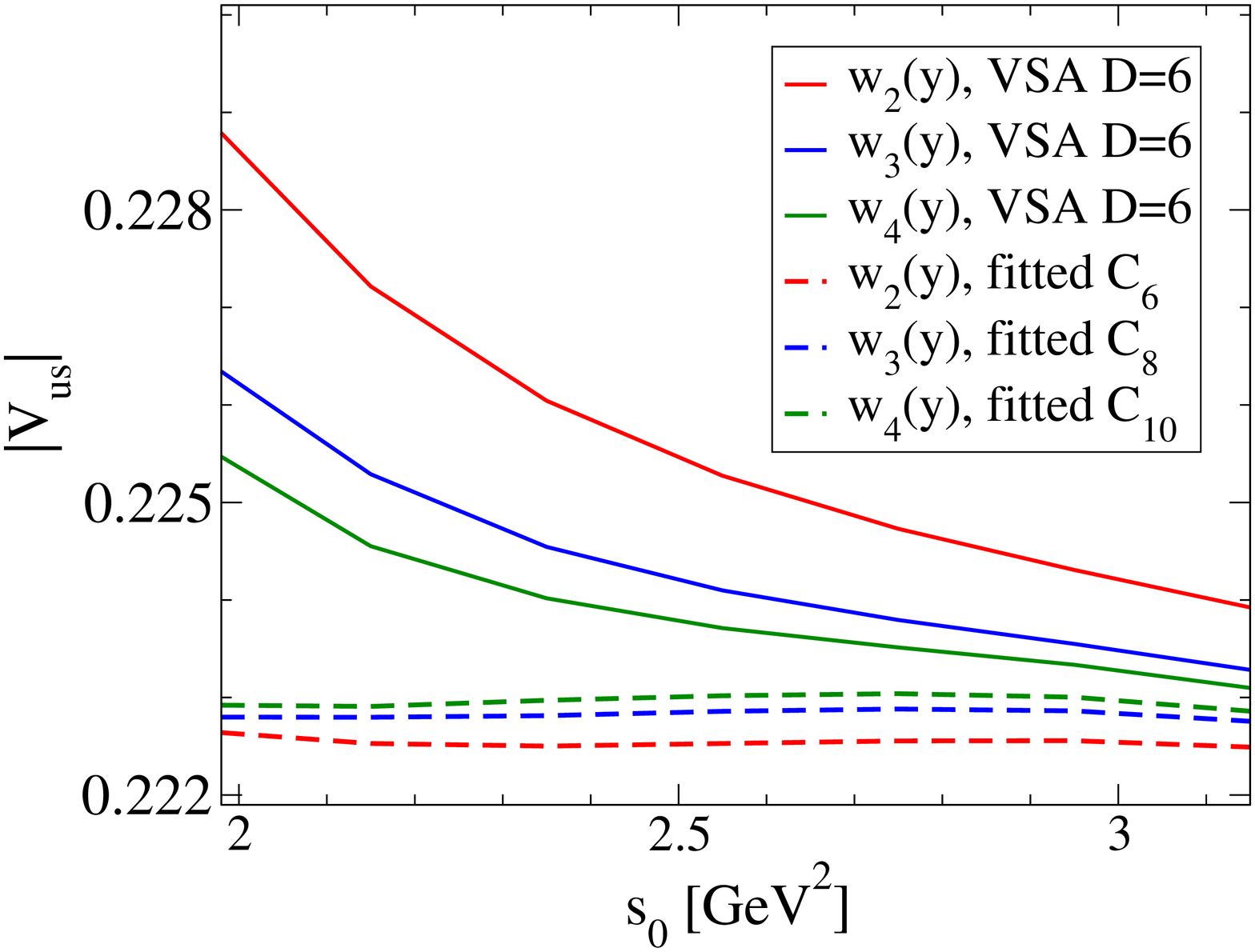}
\caption{Left panel: $\vert V_{us}\vert$ obtained from the $w_\tau$
and $\hat{w}$ FESRs using the conventional implementation~\cite{gamizetal}
OPE treatment, including use of the CIPT prescription for the $D=2$
series. Right panel: Comparison of the conventional implementation results
for $\vert V_{us}\vert$ from the $w_{2,3,4}$ FESRs with those obtained
using the central fitted values of $C_{6,8,10}$, now using the
FOPT $D=2$ prescription favored by lattice results.}
\label{fig1}
\end{figure}

The conventional implementation of Eq.~(\ref{tauvussolution})~\cite{gamizetal}
employs $w=w_\tau$ and $s_0=m_\tau^2$. This has the advantage that the
spectral integrals $R^{w_\tau}_{V+A;ud,us}(m_\tau^2)$ can be determined 
using only the inclusive non-strange and strange hadronic $\tau$ branching 
fractions, but the disadvantage that assumptions have to be made about the 
higher dimension $D=6,8$ OPE contributions in priniciple present for a 
degree $3$ weight like $w_\tau$. The restriction to a single $w$ and single 
$s_0$ precludes subjecting these assumptions to $w$- and $s_0$-independence 
self-consistency tests. It is this conventional implementation which leads 
to the long-standing puzzle of inclusive $\tau$ $\vert V_{us}\vert$ 
determinations $>3\sigma$ low relative the 3-family-unitarity expectations
($\vert V_{us}\vert =0.2258(9)$ for the $\vert V_{ud}\vert$ of
Ref.~\cite{ht14}), the most recent such determination,
$\vert V_{us}\vert = 0.2176(21)$~\cite{tauvusckm14}, e.g., lying
$3.6\sigma$ low. Tests of the conventional implementation
performed using variable $s_0$ and alternate weight 
choices~\cite{kmcwvus}, however, show sizeable $s_0$- and 
$w$-dependence~\cite{kmcwvus} (see, e.g., the left panel, and
solid lines in the right panel, of Fig.~\ref{fig1}), 
indicating systematic problems with at least some aspects 
of the conventional implementation. The dashed lines in
the right panel show the results of an alternate implementation
to be discussed below.

Two obvious potential sources exist for these instabilities. The first lies 
in the treatment of $D=6$, $8$ OPE contributions. In both the conventional 
implementation and generalized versions just mentioned~\cite{kmcwvus}, 
$D=6$ contributions are estimated using the vacuum saturation approximation 
(VSA), and $D=8$ contributions neglected. The resulting $D=6$ estimate is 
very small due to significant cancellations, first in the individual 
$ud$ and $us$ V+A sums and, second, in the FB difference of these sums. 
Such strong cancellations make use of the VSA estimate potentially 
dangerous, given the sizeable, channel-dependent VSA breaking observed in the 
flavor $ud$ V and A channels~\cite{dv7}. The second possibility 
concerns the slow convergence, at the correlator level, of the $D=2$ 
OPE series. With $\bar{a}=\alpha_s(Q^2)/\pi$, 
and $\alpha_s(Q^2)$, $m_s(Q^2)$ the running coupling and strange quark
mass in the $\overline{MS}$ scheme, one has, to four loops~\cite{bckd2ope}
(neglecting $O(m^2_{u,d}/m^2_s)$ corrections)
\begin{eqnarray}
&&\left[\Delta\Pi_\tau (Q^2)\right]^{OPE}_{D=2}\, =\, {\frac{3}{2\pi^2}}\,
{\frac{m_s(Q^2)}{Q^2}} \left[ 1\, +\, {\frac{7}{3}} \bar{a}\,
+\, 19.93 \bar{a}\,^2 \, +\, 208.75 \bar{a}\,^3\, +\cdots\ \right]\ .
\label{d2form}\end{eqnarray}
With $\bar{a}(m_\tau^2)\simeq 0.1$, convergence at the spacelike point on
the contour $\vert s\vert = s_0$ is marginal at best. This raises the
question of truncation order and truncation error estimates for the
corresponding integrated series. The $D=2$ convergence issue also shows 
up in the significant difference (increasing from $\sim 0.0010$ to 
$\sim 0.0020$ between 3- and 5-loop truncation order) in $\vert V_{us}\vert$ 
results obtained using alternate (fixed-order (FOPT) and contour-improved 
(CIPT)) prescriptions for the truncated integrated $D=2$ series 
which differ only by terms beyond the common truncation order~\cite{kmcwvus}.

In what follows, we first investigate the treatment of the $D=2$ OPE 
series using lattice data for $\Delta\Pi_\tau$, then test the 
$D=6,\, 8$ assumptions of the conventional implementation by comparing 
FESR results for a judiciously chosen pair of weights, $w_\tau (y)$ and 
$\hat{w}(y)=(1-y)^3$, $y=s/s_0$. We then present results obtained 
employing an alternate implementation of the FB FESR approach
suggested by these investigations.

\section{Lattice and continuum investigations of the OPE representation of 
$\Delta\Pi_\tau$}

Data for $\Delta\Pi_\tau (Q^2)$ over a wide range of Euclidean $Q^2$
can be generated using the lattice, with an appropriate cylinder cut 
applied to avoid lattice artifacts at high $Q^2$. This issue has been 
studied in detail for the ensemble employed here in a recent analysis 
focused on determining $\alpha_s$ from lattice current-current 
two-point function data~\cite{hlms15}. Here we first consider data at 
$Q^2$ high enough that $\left[\Delta\Pi_\tau\right]_{OPE}$ will be
safely dominated by its leading $D=2$ and $4$ contributions. $D=4$ 
contributions are determined by light and strange quark masses and 
condensates and hence known. We take FLAG results for the physical quark 
masses~\cite{FLAG}, the light condensate from GMOR, and the strange 
condensate from $\langle \bar{s}s\rangle /\langle \bar{u}u\rangle$.
The latter is determined using the HPQCD physical-$m_q$ version of this 
ratio~\cite{hpqcdcondratio}, translated to the $m_q$ of the 
ensemble employed using NLO ChPT~\cite{gl85}. We then consider
various combinations of truncation order and schemes for resumming logs 
for the $D=2$ OPE series, investigating whether any of these choices
produce a good match between the resulting $D=2+4$ OPE sum and 
lattice data in the high-$Q^2$ region.

\begin{figure}[ht]
\includegraphics[width=.38\textwidth,angle=270]
{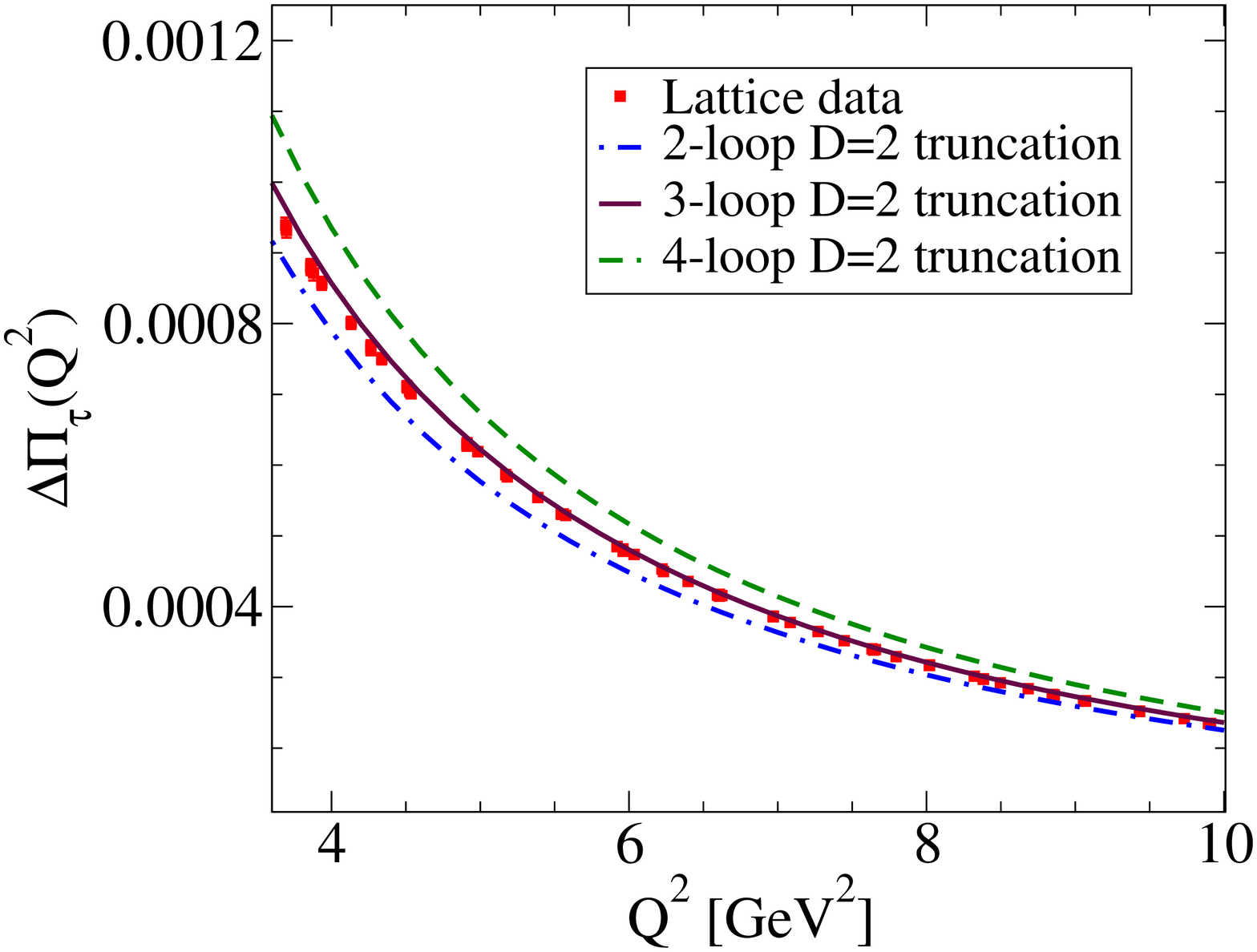}
\ 
\includegraphics[width=.38\textwidth,angle=270]
{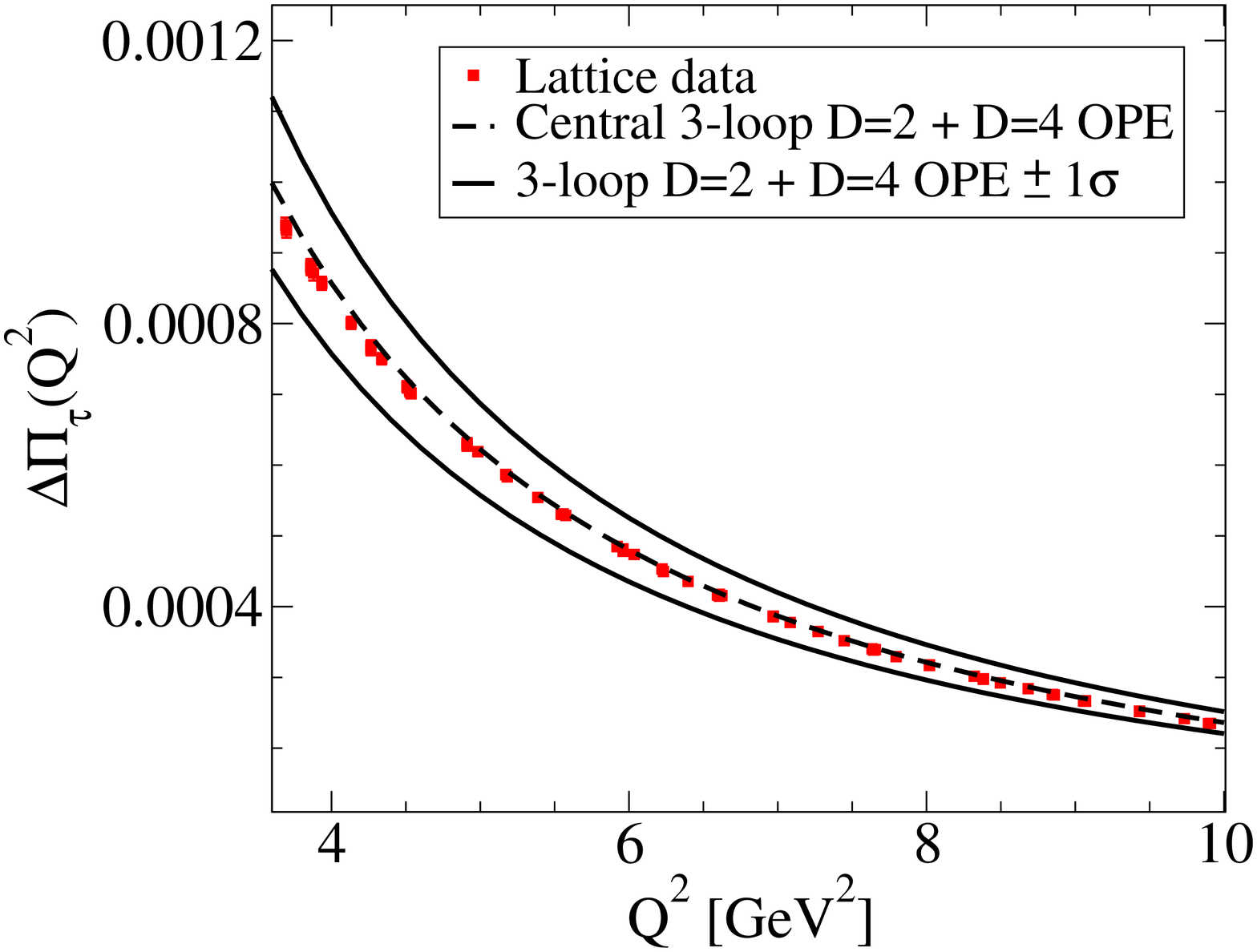}
\caption{Left panel: comparison of lattice data and OPE $D=2+4$ expectations
for the various truncation orders and the fixed-scale treatment
of the $D=2$ series. Right panel: lattice data and the $D=2+4$ OPE sum, 
with conventional OPE error estimates, for the 3-loop-truncated, fixed-scale 
$D=2$ treatment}
\label{fig2}
\end{figure}

For this high-$Q^2$ study, we employ the RBC/UKQCD $n_f=2+1$, 
$32^3\times 64$, $1/a=2.38$ GeV, $m_\pi\sim 300$ MeV domain wall 
fermion ensemble~\cite{rbcukqcdfine11}. We find (see the left
panel of Fig.~\ref{fig2}) that 3-loop $D=2$ truncation with 
fixed-scale (the analogue of the FOPT FESR prescription) provides 
an excellent OPE-lattice match over a wide range of $Q^2$, extending 
from $\sim 10$ GeV$^2$ down to $\sim 4$ GeV$^2$. The fixed-scale choice 
is, moreover, superior to the alternate local-scale ($\mu^2=Q^2$) 
choice (analogous to the CIPT FESR prescription). The right panel of
Fig.\ref{fig2} shows the $D=2+4$ OPE error band obtained using 
the 3-loop-truncated, fixed-scale $D=2$ OPE treatment and conventional 
OPE error estimate methods. The resulting, nominally naive error 
turns out to be, in fact, extremely conservative, despite the very 
slow convergence of the $D=2$ series. Clear evidence (to be detailed
elsewhere) also exists for the onset, below $Q^2\sim 4$ GeV$^2$, of 
significantly larger $D>4$ contributions than expected based on 
the VSA $D=6$ condensate employed in the conventional implementation. 
With no means of selectively isolating contributions of different $D>4$
in the Euclidean lattice data, further investigation of the higher $D$ 
question requires continuum FESR methods.

For our continuum FESR studies, we employ the $D=2$ and $4$ OPE 
treatment favored by lattice data, detailed above. As input on the 
spectral integral sides, we employ $\pi_{\mu2}$, $K_{\mu 2}$ 
and Standard Model expectations for the $\pi$ and $K$ pole 
contributions, recent ALEPH data for the continuum $ud$ V+A 
distribution~\cite{aleph13}, BaBar~\cite{babarkmpi0} and 
Belle~\cite{bellekspi} results for the $K^-\pi^0$ and $\bar{K}^0\pi^-$
distributions, BaBar results~\cite{babarkpipiallchg} for the
$K^-\pi^+\pi^-$ distribution, Belle results~\cite{bellekspipi} for the 
$\bar{K}^0\pi^-\pi^0$ distribution and 1999 ALEPH 
results~\cite{alephus99} for the ``residual'' distribution 
involving strange modes not remeasured by the B-factory experiments. 
The unit-normalized BaBar and Belle exclusive mode distributions 
must be normalized using experimental branching fractions. In the 
results quoted below we employ HFAG strange exclusive mode 
branching fractions, with the exception of $K^-\pi^0$, for which we 
employ the update contained in the recent BaBar Adametz 
thesis~\cite{adametzthesis}, performing 
an accompanying very small rescaling of the continuum $ud$ V+A 
distribution to restore unitarity.

Neglecting $\alpha_s$-suppressed logarithmic corrections, the $D>4$ OPE 
contributions to $\Delta\Pi_\tau (Q^2)$ can be written $\sum_{D> 4} C_D/Q^D$ 
with $C_D$ effective condensates of dimension $D$. The degree 3 weights 
$w_\tau (y)=1-3y^2+2y^3$ and $\hat{w}(y)=1-3y+3y^2-y^3$ generate integrated 
OPE contributions up to $D=8$. The associated $D>4$ contributions, 
\begin{equation}
-{\frac{3C_6}{s_0^2}}\, -\, {\frac{2C_8}{s_0^3}}\ \ {\rm for\ }w_\tau 
\qquad {\rm and}\qquad 
{\frac{3C_6}{s_0^2}}\, +\, {\frac{C_8}{s_0^3}}\ \ {\rm for\ }\hat{w},
\label{wtauwhatcomp}\end{equation}
differ in sign, with the two $D=6$ contributions identical in magnitude
and the magnitude of the $\hat{w}$ $D=8$ contribution half that of $w_\tau$.
It follows that, if the assumptions of the conventional implementation
are correct and $D=6,\, 8$ contributions are basically negligible in the
$w_\tau$ FESR, this will necessarily also be the case for the $\hat{w}$ FESR.
The $\vert V_{us}\vert$ results obtained from the two FESRs should 
then agree and, moreover, be $s_0$-independent. In contrast, 
if the $D=6$ and/or $8$ contributions to the $w_\tau$ FESR are 
not, in fact, negligible, one should see $s_0$-instabilities of opposite 
sign, decreasing in magnitude with increasing $s_0$, for the results 
of $\vert V_{us}\vert$ obtained from the two FESRs. The left panel
of Fig.~\ref{fig1} shows that it is the latter scenario which is 
realized. The sizeable $s_0$- and weight-choice dependences demonstrate 
unambiguously that the assumptions underlying the conventional
implementation are untenable, and that the $3\sigma$ low $\vert V_{us}\vert$
results obtained employing them are afflicted with significant 
previously unquantified systematic uncertainties.

\section{An alternate implementation of the FB FESR approach}
With previously employed methods for estimating $D>4$ effective 
OPE condensates shown to be unreliable, one has no option but to fit these
condensates to data. This requires working with FESRs 
involving variable $s_0$ and hence precludes determining the required 
spectral integrals solely in terms of inclusive hadronic branching 
fractions. To suppress possible duality violating contributions, we 
restrict our attention to FESRs with weights having at least a double 
zero at $s=s_0$. The weights $w_N(y)=1-{\frac{N}{N-1}}y+{\frac{1}{N-1}}y^N$, 
$N\ge 2$~\cite{my08} are particularly convenient since 
they yield a single $D>4$ OPE contribution (with $D=2N+2$). With $D=2+4$ 
OPE contributions under control, as discussed above, this leaves 
$\vert V_{us}\vert$ and $C_{2N+2}$ as the only parameters to be fit to the 
$w_N$-weighted, $s_0$-dependent spectral integrals. Further tests of the 
analysis are provided by checking that (i) the $\vert V_{us}\vert$ obtained 
from the different $w_N$ FESRs are in agreement and (ii) the fitted $C_D$ 
are physically sensible (i.e., show FB cancellation relative to the results 
of Ref.~\cite{dv7} for the corresponding flavor $ud$ condensates). We 
have analyzed the $w_N$ FESRs for $N=2,3,4$ and verified that the results 
pass these self-consistency tests. In the right panel of Fig.~\ref{fig1} 
we display the results obtained by taking the central values for the 
$C_6$, $C_8$ and $C_{10}$ obtained in this analysis as input and 
solving Eq.~(\ref{tauvussolution}) for $\vert V_{us}\vert$, as a 
function of $s_0$, for each of the $w_2$, $w_3$ and $w_4$ FESRs. The 
figure illustrates (i) the underlying excellent
match between the fitted OPE and spectral integral sets, (ii) the
excellent agreement between the results of the different FESR
analyses and (iii) the dramatic decrease in $s_0$- and weight-dependence
produced by using $D>4$ OPE contributions fit to data in place of
those based on the assumptions of the conventional implementation.
In addition, one sees that, as expected, the fitted $\vert V_{us}\vert$
lie between the $s_0$-unstable results produced by the conventional
implementation of the $w_\tau$ and $\hat{w}$ FESRs, and are
$\sim 0.0020$ higher than the results of the conventional $w_\tau$
implementation. 

%\section{Results and a comparison to other methods}
With the results from the different $w_N$ FESRs showing good
compatibility and $s_0$-stability, our final result for
$\vert V_{us}\vert$ is obtained by performing a combined 
fit to the $w_2$, $w_3$ and $w_4$ FESRs. We find{\footnote{An
analogous analysis with $K\pi$ normalization not employing
the $B[K^-\pi^0\nu_\tau ]$ update of Ref.~\cite{adametzthesis}
yields $\vert V_{us}\vert = 0.2200(23)_{exp}(5)_{th}$. Of the $0.0024$
difference between this result and the conventional implementation
result~\cite{tauvusckm14} noted above, $0.0005$ results
from the use of $K_{\mu 2}$ for the $K$ pole contribution; the
remainder is due to presence of the $D=6,\, 8$ contributions 
not correctly accounted for by the assumptions of the conventional 
implementation. Note that the normalization of the two-mode
$K\pi$ sum produced by the $B[K^-\pi^0\nu_\tau ]$ update is
in good agreement with the results of the dispersive
study of $K\pi$ detailed in Ref.~\cite{aclp13}.}}
\begin{equation}
\vert V_{us}\vert = 0.2228(23)_{exp}(5)_{th}\ ,
\label{adametzvusresult}\end{equation}
in excellent agreement with the results, $0.2235(4)_{exp}(9)_{th}$ and
$0.2231(4)_{exp}(7)_{th}$, obtained using the 2014 FlaviaNet experimental 
$K_{\ell 3}$ update~\cite{kell3andKratiosvus} and most recent
$n_f=2+1$~\cite{rbcukqcdfplus0} and $n_f=2+1+1$~\cite{fnalmilcfplus0} 
lattice results for $f_+(0)$. It is also compatible within 
errors with (i) the results, $0.2251(3)_{exp}(9)_{th}$ and 
$0.02250(3)_{exp}(7)_{th}$ obtained using the 2014 experimental
$\Gamma [K_{\mu 2}]/\Gamma [\pi_{\mu 2}]$ update~\cite{kell3andKratiosvus}
and most recent $n_f=2+1$~\cite{rbcukqcdfkoverfpi} and 
$n_f=2+1+1$~\cite{fnalmilcfkoverfpi} lattice determinations of 
$f_K/f_\pi$ and (ii) the expectations of 3-family unitarity.
It is worth noting that, among these methods, the one having the smallest 
theory error is the FB FESR determination, which error, as we have seen, is 
very conservative. At present the experimental error on
the FB FESR determination (resulting almost entirely from 
uncertainties in the $us$ exclusive mode distributions) is larger
than those of the competing methods, but
this error is currently dominated by the uncertainty on the
branching fraction normalizations for the exclusive strange modes,
and systematically improvable in the near future.
%\begin{table}
%\begin{tabular}{...}
%...
%\end{tabular}
%\caption{This is the caption of the table.}
%\label{tab1}
%\end{table}

\end{document}